# Long spin relaxation times in a transition metal atom in direct contact to a metal substrate

**J. Hermenau[1], M. Ternes[2], M. Steinbrecher[1], R. Wiesendanger[1] and J. Wiebe[1,*]**

[1] Department of Physics, Hamburg University, D-20355 Hamburg, Germany

[2] Max-Planck Institute for Solid State Research, D-70569 Stuttgart, Germany

*Corresponding author: jwiebe@physnet.uni-hamburg.de

**Long spin relaxation times are a prerequisite for the use of spins in data storage or nanospintronics technologies. An atomic-scale solid-state realization of such a system is the spin of a transition metal atom adsorbed on a suitable substrate. For the case of a metallic substrate, which enables directly addressing the spin by conduction electrons, the experimentally measured lifetimes reported to date are on the order of only hundreds of femtoseconds. Here, we show that the spin states of iron atoms adsorbed directly on a conductive platinum substrate have an astonishingly long spin relaxation time in the nanosecond regime, which is comparable to that of a transition metal atom decoupled from the substrate electrons by a thin decoupling layer. The combination of long spin relaxation times and strong coupling to conduction electrons implies the possibility to use flexible coupling schemes in order to process the spin-information.**

The high sensitivity of magnetic quantum systems to their closest environment can be helpful in quantum sensors in order to improve their performance in comparison to classical detectors [1] or disturbing when large spin relaxation times (typically labeled with $T_1$) and/or spin coherence times (typically referred to as $T_2$) are required. Robust magnetic properties are especially desired for quantum computing applications [2], which are based on the quantum mechanical superposition of states and thus require long $T_1$ and $T_2$ times, as well as for classical storage and spintronics applications, whose functionality particularly depends on long spin relaxation times $T_1$. If the spin is coupled to a conductive substrate for the purpose of writing and reading the spin state, spin-flip scattering of the conduction electrons typically limits both time scales drastically. Therefore, in the field of atomic spintronics and data storage, the search for long spin-lifetimes of atomic-sized magnets was mainly concentrated on systems which are decoupled from conduction electrons. In these systems, the magnetic moment bearing orbitals were protected against external perturbations by the use of insulating interlayers [3-5] and/or rare-earth elements [6-8] or molecular ligand fields [9]. Also the effects of superconducting [10] or semiconducting [11] substrates were investigated. Due to the suppressed spin-flip scattering in such systems, long $T_1$ times of transition metal atoms up to milliseconds have been realized, as revealed by pump-probe measurements and real-time observations [3].

On the other hand, a strong decoupling of the magnetic moment hinders to access the contained magnetic information, which is a drawback relating to individual reading and writing [12], and also to controllably couple several spins, which is required for processing of the spin information [13]. In this respect, adsorption of transition metal atoms directly on heavy substrate materials offers a high tunability of the couplings via distance dependent collinear [13] or non-collinear Ruderman-Kittel-Kasuya-Yosida (RKKY) interactions [14, 15]. However, for individual transition metal atoms adsorbed on metal substrates, very short lifetimes in the regime of hundreds of femtoseconds were derived from the broadening of the excitation steps in inelastic scanning tunneling spectroscopy (ISTS) [16-19]. Since $T_2$ is typically orders of magnitudes shorter than $T_1$ [20], the lifetime broadening in ISTS is most probably dominated by the spin coherence time. A direct measurement of $T_1$ and $T_2$ in these systems was missing so far.

Here, we show that the spin relaxation time $T_1$ of an iron (Fe) atom in direct contact to the heavy substrate material platinum (Pt) is on the order of a nanosecond and thereby comparable to the spin relaxation times that have been found for transition metal atoms on thin insulating layers. All metal based spin processing schemes using non-collinear substrate electron mediated interactions between neighboring single transition metal atoms thereby get within reach [14, 15, 21].

## Results

**Comparison of spin-averaging and spin-resolved ISTS** A typical area of the prepared Pt(111) surface with randomly distributed Fe atoms is shown in the constant-current scanning tunneling microscope (STM) image in Fig.1a. On platinum the Fe atoms can be adsorbed on the two different hollow sites, face centered cubic (fcc) or hexagonally closed packed (hcp), of the (111) lattice. The type of adsorption site can be identified from the characteristic spin-excitation energies detectable in spin-averaging inelastic scanning tunneling spectroscopy (ISTS) [17]. ISTS on fcc Fe atoms (Fig.1b, black curve) reveals a falling and a rising step at $-0.75\,\mathrm{mV}$ and $+0.75\,\mathrm{mV}$, respectively, with equal step heights, corresponding to the energy that it takes to excite the spin out of the ground state, with the spin oriented maximally perpendicular to the surface (up or down), into the first excited state [17]. The spin-averaging ISTS data could be reproduced using an effective spin Hamiltonian $\widehat{H} = g\mu_B \mathbf{B}\widehat{\mathbf{S}} + D\widehat{S}_z^2$ with a spin of $S = \frac{5}{2}$, a uniaxial magnetic anisotropy constant of $D = -0.19\,\text{meV}$ and a Landé $g$-factor of $g = 2.4$ [17]. In this model, the relatively large broadening of the steps was interpreted as due to lifetime broadening, revealing lifetimes of only $0.7\,\text{ps}$ [17]. The easy-plane anisotropy of the hcp Fe atom with $D > 0$ leads to zero energy Kondo scattering between the two degenerate ground states with $S_z = \pm\frac{1}{2}$ [22] thereby limiting the ground state lifetime. Therefore, we will focus on the out-of-plane system of fcc Fe here.

A drastic change of the ISTS data is observed when the tip possesses a finite and stable magnetization with a significant out-of-plane component (Supplementary Note 1). As shown for three different fcc atoms marked in Fig.1a, the corresponding spin-polarized (SP-)ISTS spectra in Fig.1b are strongly asymmetric with respect to $V = 0\,\text{V}$, i.e. the heights of the steps at $\pm 0.75\,\text{mV}$ and the conductance levels at larger bias are different on the positive and negative bias side, with an inverted step slope at negative bias. Moreover, the steps reveal a peak-like overshooting. Additionally, without changing these main characteristics, the spectra recorded on different atoms differ significantly. This must be attributed to a high sensitivity of the underlying processes on the local environment of the particular atom.

In an externally applied out-of-plane magnetic field $B_{\text{ext}}$ (Fig.2, see data with additional values of $B_{\text{ext}}$ and $I_s$ in Supplementary Fig.2), the step-like features in the SP-ISTS spectra shift towards higher bias according to the Zeeman interaction [17]. Most importantly, they strongly change their appearance and asymmetries with respect to $V = 0\,\text{V}$ when the field polarity is reversed. The spectra at negative $B_{\text{ext}}$ show steps with asymmetric heights and an overshooting characteristics, while the spectra at positive $B_{\text{ext}}$ reveal steps with inverted slope. Note, that the data was taken using the same magnetic tip with a sensitivity to the out-of-plane component of the magnetization, which is not affected by $B_{\text{ext}}$. Similar shapes of spectra and asymmetries in the step heights with respect to $V = 0\,\text{V}$ are well-known characteristics of non-equilibrium effects in SP-ISTS of an atomic spin in the presence of a magnetic field [23, 24]. Our observation of a strongly asymmetric appearance of the SP-ISTS spectra in zero magnetic field (Fig.1b) further points towards an additional stabilization of the atom's magnetization by the magnetic tip as we will show below.

The peak-like overshooting of the excitation steps can principally arise from higher order Kondo scattering effects [25, 26] or from pumping processes enabled by long spin relaxation times of the excited states [4, 24, 27]. In fact, the characteristic features in the spectra show a systematic dependence on the stabilization current $I_{\text{stab}}$, in particular for small negative $B_{\text{ext}}$ (Fig.2). This strongly indicates that dynamic processes are the origin of the complex line shapes [5, 28]. In general, dynamic processes are only relevant in ISTS if $T_1$ of the excited states is at least on the order of the average time between two tunneling electrons $\tau_I(V) = \frac{e}{I(V)}$, where $e$ is the elementary charge and $I(V)$ the measured tunneling current. Even at the lowest stabilization current $I_{\text{stab}} = 500\ pA$ used, the line shapes still indicate dynamic processes. A lower limit of $T_1$ can hence be estimated to about $\tau_{\min}(V=0) = 320\ \text{ps}$. This value is three orders of magnitude larger than the time that has been estimated from the lifetime broadening of the excitation steps [17].

**Spin relaxation times** In order to extract the spin relaxation times from the SP-ISTS spectra quantitatively, the experimental data is simulated within a third order perturbation theory model [25] (see Fig.3, data and simulations for further values of $I_{\text{stab}}$ are given in Supplementary Figs.3 and 4). It contains the coupling of substrate and tunneling electrons to the atom's spin, which is described by an effective spin Hamiltonian (see Methods) that takes into account the Zeeman energy and the complete set of anisotropy constants for the $S = \frac{5}{2}$ system in a crystal field of $C_{3v}$-symmetry, i.e. two axial anisotropy constants $B_2^0$ and $B_4^0$ as well as a transverse anisotropy constant $B_4^3$ [29]. The latter anisotropy leads to a mixing of three subsets of the $\hat{S}_z$ eigenstates (Fig.3e) into the six eigenstates $\Psi_i$ of the Hamiltonian, whose energy levels in a positive magnetic field are illustrated in Fig.3f for the present case of out-of-plane anisotropy ($B_2^0 < 0$) [17]. The model consistently reproduces almost every detail of the discussed line shapes for the complete set of measured SP-ISTS spectra using different magnetic fields and stabilization currents (Fig.3a and Supplementary Figs.3, 4). To this end, the model parameters have been slightly varied within reasonable constraints (see Table 1, Supplementary Tables 1 and 2, and Supplementary Fig.5), and we will later discuss the resulting offset magnetic field and a systematic dependence of the electron-to-local-spin coupling constants which are caused by tip-related effects. As a side note, we would like to mention that the SP-ISTS spectra taken at $B_{\text{ext}} = +0.1\text{T}$ closely resemble the *ab-initio* calculated renormalized local vacuum density of states from Ref. [30].

Justified by the excellent agreement between simulation and data, we use the model to extract the lifetimes $\tau_i$ and occupation probabilities $p_i$ of the six spin-states $\Psi_i$ in dependence of the bias voltage (Fig.3b-d). Note, that, with a spin larger than $S = \frac{1}{2}$, an individual spin relaxation time $\tau_i$ ($i > 2$) is assigned to each excited state $\Psi_i$ [31], which is the average time before any scattering event transfers the system into a different eigenstate. In Fig.3b,c the bias voltage dependence of $\tau_i$, calculated from the simulations of the $B_{\text{ext}} = 0\,\text{T}$ data recorded with $I_{\text{stab}} = 1\,\text{nA}$, are plotted together with the average time between two tunneling electrons $\tau_I(V) = \frac{e}{I(V)}$. Above the voltage of the first spin-excitation (see the vertical lines at $\pm 0.75\,\text{mV}$), all six states have a lifetime considerably longer ($\Psi_1$, $\Psi_2$, $\Psi_3$, $\Psi_4$) than, or on the same order ($\Psi_5$, $\Psi_6$) of $\tau_I(V)$, corroborating the spin-pumping effect of the tunneling electrons in this system. In particular, the spin relaxation times of the first excited states $\Psi_3$ and $\Psi_4$ for a bias voltage close to the excitation threshold are longer than $1\,\text{ns}$, which is three orders of magnitude larger than the lifetime determined from the analysis of the broadening in ISTS measurements [17]. Surprisingly long lifetimes are found for the two ground states ($\Psi_1$, $\Psi_2$) as shown in Fig.3b. For voltages above the tunneling electron excitation threshold the lifetimes of both states decrease due to the interaction with an increasing number of tunneling electrons, as shown by the nearly constant ratio $\frac{\tau_{1/2}}{\tau_I(V)} \sim 10$. This means that on average every tenth tunneling electron excites the spin out of the ground state. Below this threshold the extracted ground state lifetimes increase up to values ranging between micro- and milliseconds.

**Spin transfer torque** Resulting from the comparatively long lifetimes of the spin states, there is a significant non-thermal occupation of the higher energy spin states due to the interplay of the spin-pumping effect and a non-vanishing spin transfer torque caused by the spin-polarization of the tunneling current [28]. Fig.3d shows the occupation probabilities $p_i$ of the states $\Psi_i$ for the situation of a small positive magnetic field, where $\Psi_1$ on the left side of the anisotropy barrier is the non-degenerate ground state (cf. Fig.3f). Note, that there is an effective magnetic field induced by the magnetic tip (see below). Consequently, for bias voltages of the tunneling electrons below the excitation threshold (vertical lines in Fig.3d) the occupation probability $p_1$ of $\Psi_1$, is highest. However, if the spin is driven by tunneling electrons at a negative bias above the excitation threshold, the situation gets reversed. Now, the occupation of the energetically higher lying states $\Psi_2$, $\Psi_4$, and $\Psi_6$ on the right side of the anisotropy barrier is larger as compared to their energetically lower partners $\Psi_1$, $\Psi_3$, and $\Psi_5$ on the left side. Obviously, for this choice of the bias polarity, subsequent excitations induced by the spin-polarized tunneling electrons pump the spin from the energetically favored ground state across the anisotropy barrier, i.e. the spin is reversed [28]. This spin-pumping effect enables the atom's spin to be driven into the desired ground state only by choosing the proper bias voltage polarity [5].

**Tip effects on the spin states and their lifetime** We finally focus on effects of the tip on the Fe atom in addition to the spin-pumping of the tunneling electrons. First, there is a strong asymmetry with respect to zero bias even at zero magnetic field (Fig.1b, Fig.2, Fig.3a), which hints towards an additional inherent magnetic field. Indeed, as shown in Fig.4a, the magnetic field $B_{\text{sim}}$ resulting in the correct simulation of the data has a small offset of about $+0.1\,\text{T}$ to the applied magnetic field $B_{\text{ext}}$, which is compensated at $B_{\text{ext}} \approx -0.1\,\text{T}$ (see inset of Fig.4a), dependent on $I_{\text{stab}}$. Obviously, there is an additional effective magnetic field caused by the tip $B_{\text{tip}}$, whose magnitude depends on the experimentally given tip-sample conductance $G_{\text{ts}} = \frac{I_{\text{stab}}}{V_{\text{stab}}}$, where $V_{\text{stab}} = 5\,\text{mV}$ is the stabilization voltage, and thereby on the tip-atom separation adjusted by $I_{\text{stab}}$. $B_{\text{tip}}$ most probably originates from a stray field and/or an exchange interaction between tip apex and atom [32, 33]. Interestingly, as visible in the inset of Fig.4a, the resulting field at the location of the atom even reverts its orientation from negative to positive polarity when $I_{\text{stab}}$ is increased from $1\,\text{nA}$ to $16\,\text{nA}$. In the data, the effect of the $I_{\text{stab}}$ dependent reversal of the magnetic field polarity can be seen for the $B_{\text{ext}} = -0.1\,\text{T}$ data in Fig.2, where the spectra at low $I_{\text{stab}}$ resemble those taken at large negative $B_{\text{ext}}$ and the spectra at larger currents those taken at positive $B_{\text{ext}}$.

A second tip-related effect is found in the strong $G_{\text{ts}}$ dependence of the coupling of the atomic spin to the sample electron bath $G_s = \frac{e^2}{h}(J_K \rho_s)^2 \cdot S(S+1)$ that can be calculated from the simulation parameter $J_K \cdot \rho_s$ (Fig. 4c). $G_s$ quantifies the scattering efficiency of substrate electrons (see the cartoon in Fig.4d) via the Kondo coupling strength $J_K$, the electronic density of states of the substrate $\rho_s$ and the spin of the atom. In the model, the tunneling current is included in the barrier transmission coefficient $T_0^2$ only. As illustrated in Fig.4b, the corresponding tip-sample conductance $G_{\text{ts}}(T_0^2, U) = \left[\frac{1}{2}S(S+1) + 2U^2\right] \cdot T_0^2$, calculated from the averaged values of the simulated potential scattering strength $U$ and $T_0^2$ in units of the quantum of conductance $G_0$, shows a one-to-one correspondence to the experimentally adjusted $G_{\text{ts}} = \frac{I_{\text{stab}}}{V_{\text{stab}}}$. This result not only confirms the validity of the values used for $U$ and $T_0^2$, but also proves that all tunneling current related effects are covered by $T_0^2$. The observed nonlinear dependence of the simulated $G_s$ on $G_{\text{ts}}$ (Fig.4c), therefore, has to originate in another effect from approaching the tip towards the atom not considered in the model. In fact, electrons from the tip can also scatter on the atom and back into the tip (Fig.4d), which further decreases the lifetime of the spin-states [3]. The efficiency $G_{\text{tt}}$ of these scattering processes increases for increasing $I_{\text{stab}}$ due to the reduced tip-sample separation. Since our model does not include this scattering separately, its significant impact on the spectral line shape is implicitly incorporated in the simulated $G_s$, explaining its unexpected dependence on $G_{ts}$.

Due to this effect, the lifetimes of the two ground states will depend on the actual tip-sample separation even for zero bias-voltage, as shown in Fig.4e. As a direct consequence of the increasing efficiency of the tip-atom-tip scattering, a monotonic decrease of $\tau_1$ and $\tau_2$ is observed for increasing $G_{ts}$. The same effect was also found for substrate decoupled Fe atoms [3]. As shown there, the monotonic behavior of $G_s$ in Fig.4c indicates that the spin-lifetimes of the Fe atoms on Pt(111) are mainly limited by the spin-flip scattering of tip electrons, even for the largest tip-sample separations. Consequently, with further decreasing $I_{stab}$, we expect even larger zero voltage spin-lifetimes until they finally approach their intrinsic values limited by substrate electron scattering.

**Discussion**

We finally discuss the reliability of and possible reasons for the observed surprisingly long lifetimes of the spin-states of Fe atoms on Pt(111). The spin relaxation times of the excited states extracted from the simulations (Fig.3c) are comparable to the lower limit $\tau_{min}(V) = 320$ ps which was roughly estimated from $I_{stab}$. Therefore, we believe that the simulation results for the spin relaxation times of the excited states are quite reliable. $\tau_3/\tau_4$, which are the longest $T_1$ times of all excited states, are thus three orders of magnitude longer than the lifetime of $0.7$ ps extracted from the broadening analysis of spin-averaging ISTS [17]. We therefore propose, that the experimentally observed broadening of the spin-averaged ISTS spectra is dominated by the spin coherence times $T_2$ of the excited states, which typically are much shorter than the $T_1$ times. Regarding the extremely long lifetimes of the ground states, the comparison of the outcomes of Ref. [25] and Ref. [5] lead to the conclusion that a large number of scattering processes do not leave their fingerprints in SP-ISTS spectra. Such hidden scattering processes might cause an overestimation of the intrinsic spin-lifetimes of the ground states calculated by the model. Therefore, the ground state lifetimes need to be confirmed by a more direct technique as, e.g., a pump-probe experiment [4].

Ground state lifetimes in the range of micro- to milliseconds are comparable to that of Fe atoms which are decoupled from substrate conduction electrons by thin insulating layers [3]. From the simulation we conclude that the reasons for our finding of long living spin-states in Fe atoms directly coupled to the conductive Pt substrate are i) the beneficially large ratio of axial and transverse anisotropy and ii) the small coupling to the substrate electron bath $G_s$. Indeed, the simulated values for $G_s$ are on the same order of magnitude as values found for manganese atoms on the insulating Cu$_2$N substrate [25]. Note, that Pt is a large-susceptibility substrate leading to a large cluster of non-collinearly magnetized Pt atoms beneath the Fe [15, 17], which will have a drastic effect on $G_s$. It is, therefore, an interesting question, whether the long living spin-states observed here in an all-metallic system are a general property of adatom/substrate systems using substrates that almost fulfill the Stoner criterion for ferromagnetism, e.g. Rh [34] or Pd [35].

## Methods

### Experimental procedures

The measurements have been performed in a low-temperature scanning tunneling microscope facility with an external magnetic field $B_{\text{ext}}$ applied perpendicular to the sample surface [36] using the same preparation as well as the same measurement conditions as described in Ref.[15]. The cleaned Pt(111) surface was coated with a fraction of a monolayer Co resulting in monolayer islands and stripes (see Supplementary Note 1) before single Fe atoms were deposited on the pre-cooled sample leading to a statistical distribution of Fe atoms on the Pt terraces. The spin-polarization of the tip was achieved by using a Cr-coated tungsten tip whose spin-contrast was optimized by dipping into Co or picking up Fe atoms until a strong and magnetically stable out-of-plane contrast was observed in the differential conductance channel on the Co islands or stripes.

SP-ISTS measurements were performed by stabilizing the tip at the bias voltage $V_{\text{stab}}$ (applied to the sample) and current $I_{\text{stab}}$ on top of the fcc Fe atoms. During the voltage sweep, the feedback loop was swiched off in order to keep the tip-sample separation constant. $\frac{dI}{dV}$ was recorded as the response to a small modulation voltage $V_{\text{mod}}$ ($f_{\text{mod}} = 4142\,\text{Hz}$) added to the bias using lock-in technique. All shown spectra were normalized via dividing by a substrate spectrum taken with the same measurement parameters in order to suppress spectral features resulting from the tip.

### Third order perturbation theory model

The experimental SP-ISTS data were modeled by employing a third order perturbation theory [25], which is based on the effective spin Hamiltonian

$$\widehat{H}_{C_{3v}} = g\mu_B \mathbf{B} \cdot \hat{\mathbf{S}} + (3B_2^0 - (30S(S+1) - 25)B_4^0) \cdot \hat{S}_z^{\,2} + 35B_4^0 \cdot \hat{S}_z^{\,4}$$
$$+ \frac{1}{2}B_4^3 \left( \hat{S}_z \left( \hat{S}_+^{\,3} + \hat{S}_-^{\,3} \right) + \left( \hat{S}_+^{\,3} + \hat{S}_-^{\,3} \right) \hat{S}_z \right)$$

(see Supplementary Note 3), which is consistent with a spin of quantum number $S = 5/2$ [17] in a crystal field of $C_{3v}$ symmetry [29]. Here, the first term is the Zeeman energy with the Landé $g$-factor ($g = 2.4$) [17], the Bohr magneton $\mu_B$, the magnetic field $\mathbf{B}$, and the vector of the spin operators $\hat{\mathbf{S}}$. The second and third terms contain the axial magnetic anisotropy with the operator of the $z$-component of the spin $\hat{S}_z$ and the axial anisotropy constants $B_2^0$ and $B_4^0$. The fourth term is the transverse magnetic anisotropy with the raising ($\hat{S}_+$) and lowering ($\hat{S}_-$) ladder operators and the corresponding transverse anisotropy constant $B_4^3$. Besides the anisotropy constants, the following model parameters have been varied during the fitting of the simulation to the experimental data (see Fig.4d): (i) the potential scattering parameter $U = U_{\text{pot}}/J_K$, which determines the ratio between potential and Kondo scattering; (ii) the Kondo scattering strength of the substrate electrons on the local spin $J_K \cdot \rho_s$; (iii) the coupling between tunneling electrons and the local spin $T_0^2$, which is determined by the tunneling current; (iv) the tip's spin polarization $\eta_t = \frac{\rho_\uparrow - \rho_\downarrow}{\rho_\uparrow + \rho_\downarrow}$, whose strength is given by the imbalance in the spin-resolved densities of states $\rho_{\uparrow/\downarrow}$; (v) the out-of-plane magnetic field $B_{\text{sim}}$; (vi) an in-plane magnetic field component $B_y$; and (vii) the effective temperature $T_{\text{eff}}$. The parameters which have been used in order to simulate the spectra are given in Table 1 and Supplementary Tables 1 and 2. Note, that the extracted $B_2^0$ is consistent with the value $\frac{D}{3}$ extracted from the simulation of the spin-averaged data [17]. The large values $U \gg 0$ display the high efficiency of elastic tunneling contributions through the atom as compared to inelastic transitions. The values for $\eta_t$ are in the typical range of spin-polarizations that were found for magnetically coated tips in SP-STS measurements. The effective temperature $T_{\text{eff}}$, which considers additional broadening due to spin coherence effects, is slightly larger than the thermodynamic temperature in the experiments ($T = 0.3$ K). The small in-plane magnetic fields $B_y$ are probably due to a transverse stray field of the tip. See a description of additional impacts of the simulation parameters on the simulated spectra in the Supplementary Materials.

**Data availability**

The authors declare that the main data supporting the findings of this study are available within the article and its Supplementary Information files. Extra data are available from the corresponding author upon reasonable request.

## Acknowledgements

We would like to thank Samir Lounis and Manuel dos Santos Dias for fruitful discussions. J.H., M.S., R.W., and J.W. acknowledge funding from the SFB668 and from the GrK1286 of the DFG. M.T. acknowledges funding via SFB767.

## Author contributions

J.H., M.S. and J.W. designed the experiments. J.H. carried out the measurements and did the analysis of the experimental data together with M.T.. J.H. and J.W. wrote the manuscript, to which all authors contributed via discussions and corrections.

## Competing financial interests

The authors declare no competing financial interests.

# Figures

**Figure 1 | Spin-resolved inelastic scanning tunneling spectroscopy of iron atoms on Pt(111). a**, Constant-current image of the sample surface showing randomly distributed individual Fe atoms (image size $16 \times 5 \text{ nm}^2$, $V = 5 \text{ mV}$, $I = 500 \text{ pA}$). **b**, SP-ISTS spectra of the fcc atoms marked with arrows in the corresponding color in **a** ($V_{\text{stab}} = 6 \text{ mV}, I_{\text{stab}} = 3 \text{ nA}, V_{\text{mod}} = 40 \text{ μV}, B = 0 \text{ T}$) in comparison to an ISTS spectrum recorded with a non-magnetic tip (black, atom not shown in **a**, $V_{\text{stab}} = 10 \text{ mV}, I_{\text{stab}} = 3 \text{ nA}, V_{\text{mod}} = 40 \text{ μV}, B = 0\text{T}$). All spectra are normalized by dividing a substrate spectrum taken with the same parameters.

**Figure 2 | Magnetic field and stabilization current dependence of SP-ISTS.** SP-ISTS spectra recorded on the fcc atom shown in the inset taken at the indicated magnetic fields using stabilization currents of $I_{\text{stab}} = 1 \text{ nA}$ (blue), $I_{\text{stab}} = 6 \text{ nA}$ (orange) and $I_{\text{stab}} = 16 \text{ nA}$ (purple). All spectra are normalized by dividing a substrate spectrum taken at the same parameters ($V_{\text{stab}} = 5 \text{ mV}$, $V_{\text{mod}} = 80 \text{ μV}$). For better visibility the spectra recorded at different magnetic fields are shifted by an artificial vertical offset.

**Figure 3 | Spin-state lifetimes and occupations from the simulation of SP-ISTS. a**, Simulations (magenta) of the experimental SP-ISTS spectra (gray) recorded using a stabilization current of $I_{\text{stab}} = 1 \text{ nA}$ for the indicated magnetic fields ($V_{\text{stab}} = 5 \text{ mV}$, $V_{\text{mod}} = 80 \text{ μV}$). For better visibility all spectra (except the $B_{\text{ext}} = 0 \text{ T}$ spectra) are shifted vertically by an artificial offset indicated by the shift in the conductance at $V_{\text{stab}} = 5 \text{ mV}$. The experimental data was normalized by dividing a substrate spectrum taken with the same parameters before it was rescaled to real units by the stabilization. **b**, **c**, **d** Bias voltage dependence of (**b**) the spin-lifetimes of the two ground states ($\tau_1$ and $\tau_2$), (**c**) the higher spin-states, and (**d**) the occupations of all spin-states (see **f** for the assignment between line colors and spin states) calculated from the simulation of the experimental data taken at $I_{\text{stab}} = 1 \text{ nA}$ and $B_{\text{exp}} = 0 \text{ T}$. The average time between two tunneling electrons $\tau_I = \frac{\text{e}}{I(V)}$ is given in black. The vertical dashed lines mark the voltage corresponding to the first excitation energy determined from ISTS measurements. **e**, **f**, Illustrations of the mixing of $\hat{S}_z$ eigenstates with the same line style (solid, dashed, wiggly lines) due to transversal magnetic anisotropy (**e**) and of the resulting energy levels of the spin states (**f**, $B > 0$) with colors used for the corresponding lines in (**b-d**).

**Figure 4 | STM tip related effects on the spin-state lifetime. a**, Out-of-plane magnetic field used for the simulations ($B_{\text{sim}}$) in dependence of the external field applied in the experiment ($B_{\text{ext}}$) for the indicated stabilization currents. The inset shows a zoom-in of the area marked with the dashed box. **b**, Tip-sample conductance $G_{\text{ts}}(T_0^2, U)$ calculated from the averaged simulation parameter $T_0^2$ and $U$ in dependence of the tip-sample conductance $G_{\text{ts}}$ used in the experiment. **c**, Substrate coupling $G_s$ calculated from the averaged values of $J_K\rho_s$ in dependence of the experimentally adjusted tip-sample conductance $G_{\text{ts}}$. **d**, Illustration of the different electron scattering processes which have an impact on the dynamics. **e**, Lifetimes of the two ground states $\tau_1$ (purple) and $\tau_2$ (orange) at $V = 0$ (and thus $I = 0$) in dependence of the tip-sample conductance $G_{ts}$ calculated from the simulations of the experimental data taken at $B_{\text{exp}} = 0 \text{ T}$.

# Tables

| $B_2^0$ (meV) | $B_4^0$ (meV) | $B_4^3$ (meV) | $U$ | $T_{\text{eff}}$ (K) | $\eta_{\text{t}}$ | $B_y$ (T) |
|---|---|---|---|---|---|---|
| $-0.084 \pm 0.003$ | $0.0014 \pm 0.0002$ | $0.0014 \pm 0.0015$ | 3.62 $\pm 0.55$ | 0.65 $\pm 0.06$ | 0.17 $\pm 0.02$ | 0.28 $\pm 0.14$ |

**Table 1 | Simulation parameters.** Averaged values of the anisotropy constants $B_2^0$, $B_4^0$, $B_{43}^3$, of the potential scattering parameter $U$, of the effective temperature $T_{\text{eff}}$, of the tip's spin-polarization $\eta_{\text{t}}$, and of the transverse magnetic field $B_y$. The values have been calculated by averaging of the parameters given in Supplementary Tables 1 and 2. The error is given by the standard deviation.

Figure 1

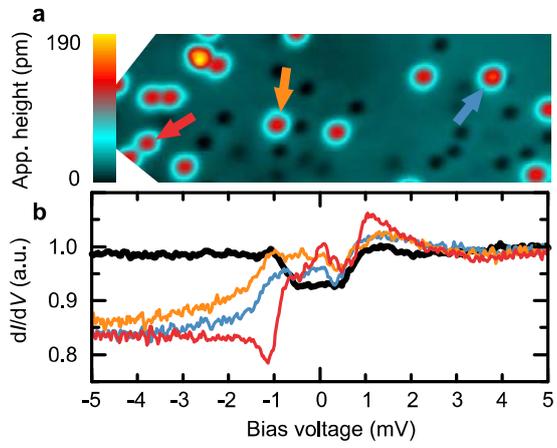

Figure 2

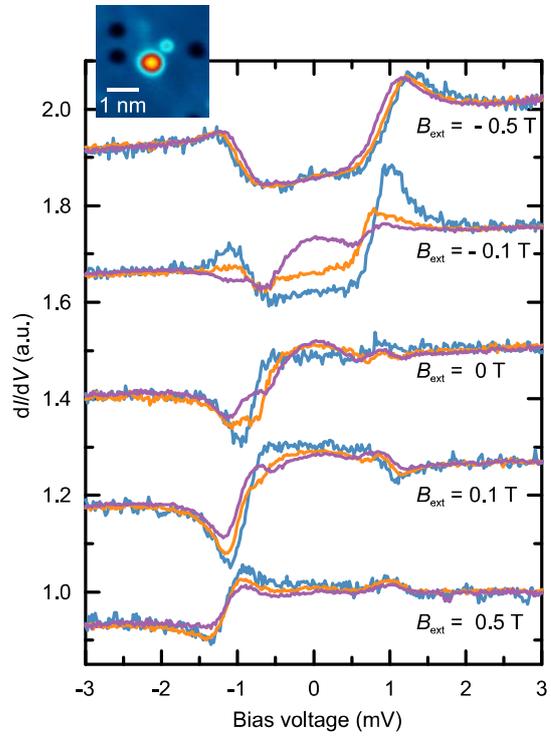

# Figure 3

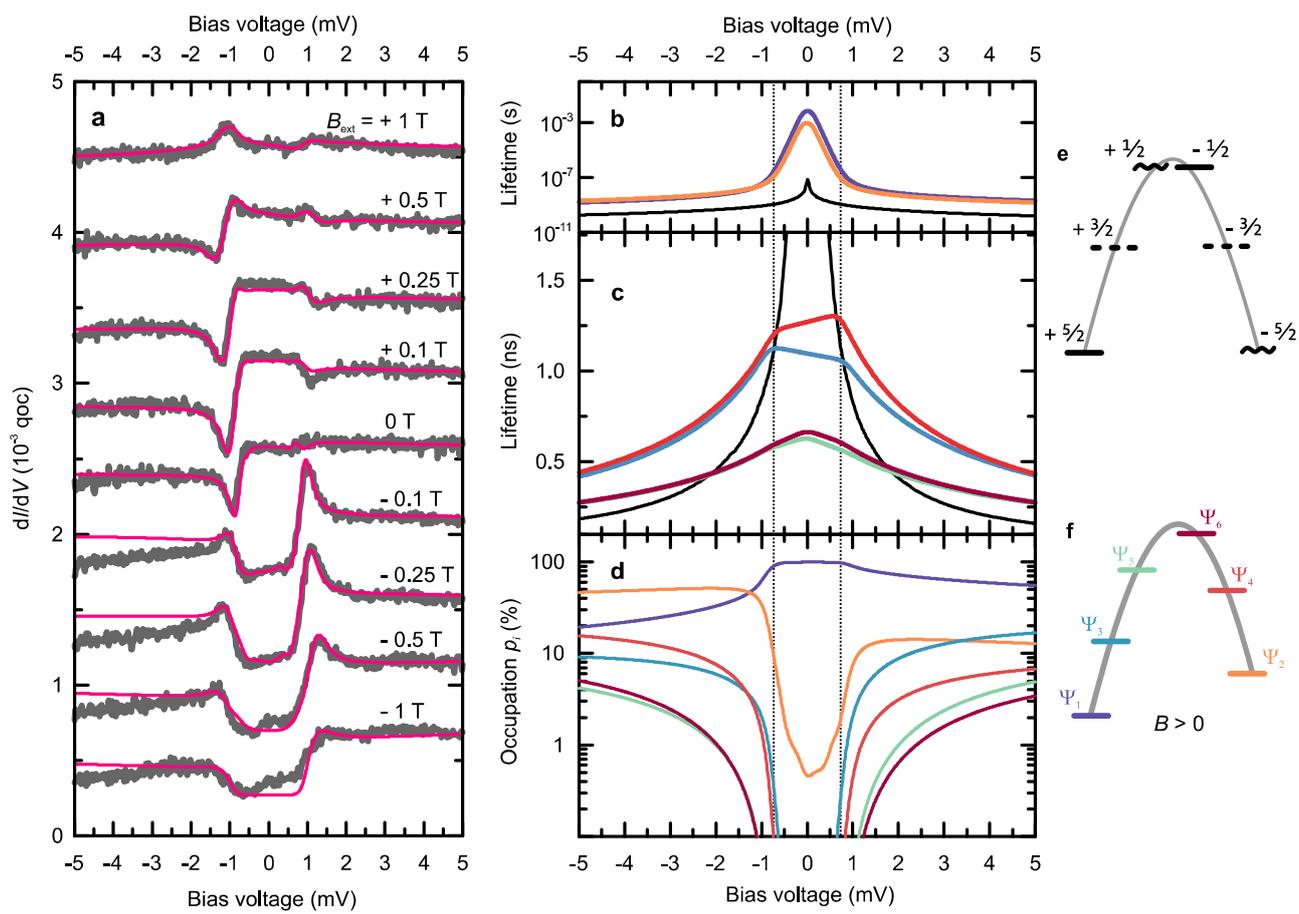

Figure 4

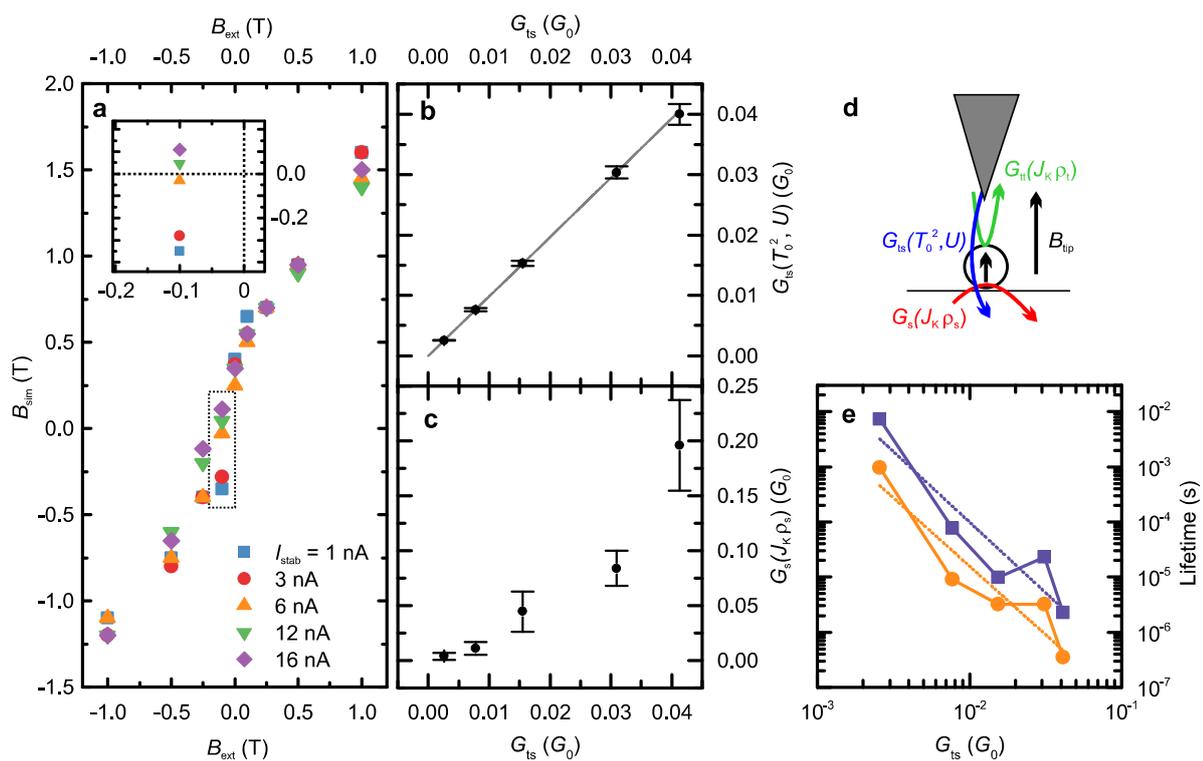

# SUPPLEMENTARY INFORMATION

**Supplementary Note 1** | **SP-ISTS on fcc Fe atoms RKKY-coupled to a Co stripe**

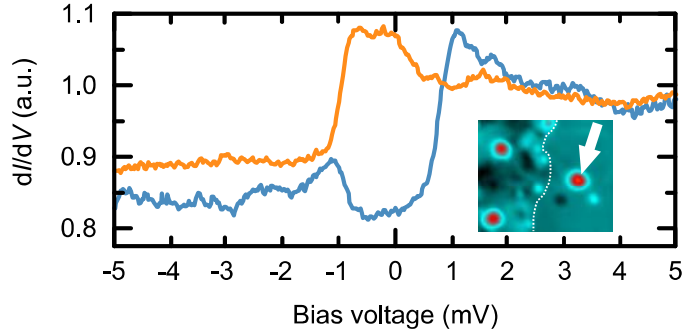

**Supplementary Figure 1** | Spin-polarized ISTS spectra recorded on the fcc Fe atom marked by the arrow in the constant-current image (inset), which is magnetically coupled to a Co monolayer stripe (see inset, left of the dashed line). The blue spectrum was taken before, and the orange spectrum after the application of a magnetic field pulse, which reversed the orientation of the out-of-plane magnetization of the Co stripe ($V_{\text{stab}} = 6\,\text{mV}$, $V_{\text{mod}} = 40\,\mu\text{V}$, $I_{\text{stab}} = 3\,\text{nA}$, $B_{\text{ext}} = 0\,\text{T}$).

Co monolayer stripes attached to the step edges and islands on the terraces of the Pt(111) substrate have been prepared as described previously[1]. They serve as a magnetically stable out-of-plane magnetized reference system for the calibration of the tip's spin-polarization[2]. All atoms investigated in the main text are well separated from the Co islands or stripes. However, we can also find Fe atoms adsorbed close to a Co stripe or island (Supplementary Fig.1), which feel a significant RKKY coupling that stabilizes the atom's spin parallel or antiparallel to the Co magnetization[2,3]. On such fcc atoms, the SP-ISTS spectra measured at $B_{\text{ext}} = 0T$ (Supplementary Fig.1) resemble those measured on the isolated atoms shown in Fig.2 of the main text at negative (blue spectrum) or positive (orange spectrum) magnetic field. When the magnetization of the nearby Co stripe is reversed by a magnetic field pulse, the spectrum changes from the blue to the yellow type. This confirms the magnetic stability of the tip in an applied magnetic field.

# Supplementary Note 2 | Magnetic field and stabilization current dependence of SP-ISTS

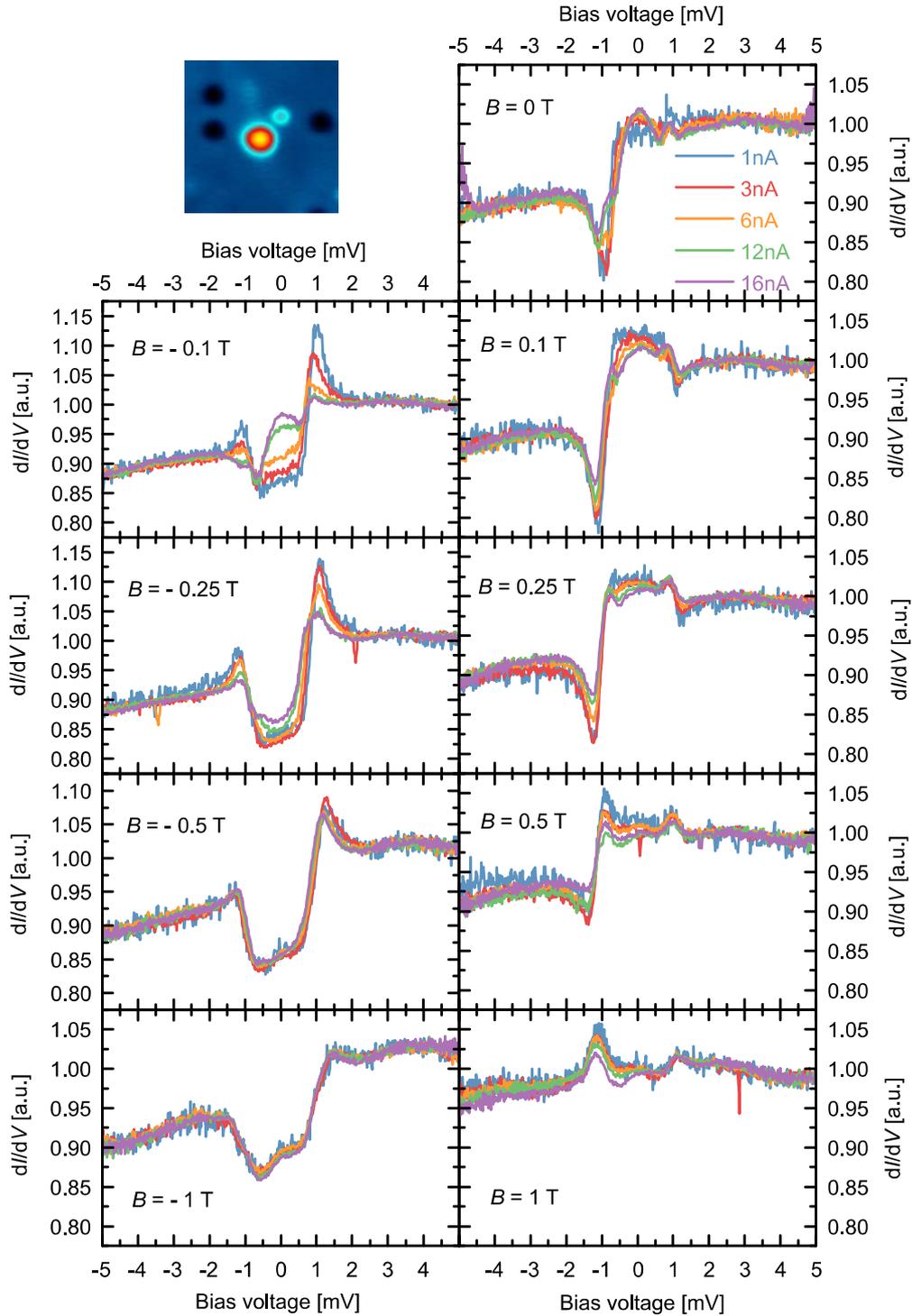

**Supplementary Figure 2** | SP-ISTS spectra recorded on the fcc atom shown in the constant-current image (inset), at the indicated magnetic fields $B_{\text{ext}}$ and stabilization currents $I_{\text{stab}}$ ($V_{\text{stab}} = 5\,\text{mV}$, $V_{\text{mod}} = 80\,\mu\text{V}$).

**Supplementary Note 3 | Simulations of the experimental data**

We simulate the electron transport between the spin-polarized tip and the Fe spin system on the Pt(111) substrate by using a perturbative scattering model in which we account for spin-flip processes in first and second order Born approximation and which has previously been used successfully on different quantum spin systems[4–8]. We divide the total Hamiltonian of the system into the Hamiltonians of the electron reservoirs of tip and substrate, between which the tunneling process via the Fe spin takes place and which, in second quantization, can be written as

$$\hat{H}_j = \sum_{\substack{j=\text{s},\text{t} \\ k,\sigma}} \epsilon_{jk\sigma} \hat{a}^\dagger_{jk\sigma} \hat{a}_{jk\sigma}, \tag{1}$$

with $\hat{a}^\dagger_{jk\sigma}$ ($\hat{a}_{jk\sigma}$) as the creation (annihilation) operators for electrons at the electrode $j = $ s (substrate) and $j = $ t (tip) with momentum $k$, spin $\sigma$, and the energy $\epsilon_{jk\sigma}$. Assuming a flat and featureless density of states in tip and substrate electrode in the small energy range of interest around the Fermi energy, we describe the reservoirs with energy independent spin density matrices $\varrho_{\text{t,s}} = |\varphi\rangle\langle\varphi| = \frac{1}{2}(\hat{\sigma}_0 + \vec{n}_{\text{t,s}} \cdot \hat{\boldsymbol{\sigma}})$. Here, the spin-polarization of the electrons from the spin-polarized tip is accounted for by the direction and amplitude of the vector $\vec{n}_\text{t}$, with $\hat{\boldsymbol{\sigma}} = (\hat{\sigma}_x, \hat{\sigma}_y, \hat{\sigma}_z)$ as the standard Pauli matrices, and $\hat{\sigma}_0$ as the $(2 \times 2)$ identity matrix. With this convention, the spin polarization is identical to the relative imbalance between majority and minority spin densities in the tip, $\eta_\text{t} = |\vec{n}_\text{t}| = \left|\frac{\rho_\uparrow - \rho_\downarrow}{\rho_\uparrow + \rho_\downarrow}\right|$, where $\uparrow$ and $\downarrow$ account for the two different spin directions along the quantization axis of the external magnetic field $\mathbf{B}_\text{ext}$ ($z$-direction). For the substrate, we assume a zero spin polarization, i. e. $\eta_\text{s} = 0$. Note, that, within this model, the induced spin-polarization of the Pt substrate is accounted for by assuming a larger Fe spin of $S = 5/2$, which results from the DFT calculated total magnetic moment of the whole cluster of Fe atom and surrounding polarized substrate atoms[9].

To model the spin-states of the Fe atom in its three-fold symmetric adsorption site on the Pt(111) substrate we utilize an effective Hamiltonian which includes all non-vanishing anisotropy constants as well as the atom's Zeeman energy:

$$\begin{aligned}\hat{H}_{\text{C}_{3v}} = & B_2^0 \cdot [3\hat{S}_z^2 - S(S+1)] \\ & + B_4^0 \cdot [35\hat{S}_z^4 - (30S(S+1) - 25)\hat{S}_z^2 + (3S^2(S+1)^2 - 6S(S+1))] \\ & + B_4^3 \cdot \tfrac{1}{2}[\hat{S}_z(\hat{S}_+^3 + \hat{S}_-^3) + (\hat{S}_+^3 + \hat{S}_-^3)\hat{S}_z] + g\mu_\text{B} \mathbf{B} \cdot \hat{\mathbf{S}} \end{aligned} \tag{2}$$

Similar as in previous publications[9] we fix the Landé-Factor to $g = 2.4$. The operator $\hat{\mathbf{S}} = (\hat{S}_x, \hat{S}_y, \hat{S}_z)$ with $\hat{S}_x = \frac{1}{2}(\hat{S}_+ + \hat{S}_-)$ and $\hat{S}_y = \frac{1}{2i}(\hat{S}_+ - \hat{S}_-)$ is the generalized total spin operator for the spin-5/2 system, $\mu_\text{B}$ is the gyromagnetic ratio, and $\mathbf{B} = \mathbf{B}_\text{ext} + \mathbf{B}_\text{tip}$ is the sum of the external magnetic field and the effective field of the tip. We permit for a small canting of the effective tip field by including, w.l.o.g., a small $y$-component of $\mathbf{B}$, i.e. $\mathbf{B} = (0, B_y, B_\text{sim})$. The magnetic anisotropy is governed by the symmetry and consist of the quadratic $B_2^0$ and cubic $B_4^0$ coefficients of the axial, out-of-plane, anisotropy. The possible threefold anisotropy governed by $B_4^3$ is also accounted for, but found to be very small with values of $B_4^3 \approx 1\,\mu\text{eV}$.

Using equations 1 and 2 we can now continue and write the total Hamiltonian as

$$\hat{H} = \hat{H}_{\text{C}_{3v}} + \hat{H}_\text{s} + \hat{H}_\text{t} + \hat{H}',  \tag{3}$$

with $\hat{H}'$ as the perturbation, which enables the tunneling of electrons from tip to substrate or vice versa via Kondo-like spin-flip or potential scattering processes and, additionally, the scattering of the bath electrons with the impurity:

$$\begin{aligned}
\hat{H}' &= \hat{V}_{s \to t} + \hat{V}_{t \to s} + \hat{V}_{s \to s} \\
\hat{V}_{t \to s} &= \sum_{\substack{\sigma,\sigma' \\ k,k'}} T_0\, \hat{a}^\dagger_{sk'\sigma'}\, \hat{a}_{tk\sigma} \left( \frac{1}{2}\hat{\boldsymbol{\sigma}} \cdot \hat{\mathbf{S}} + U\hat{\sigma}_0 \right), \\
\hat{V}_{s \to t} &= \sum_{\substack{\sigma,\sigma' \\ k,k'}} T_0\, \hat{a}^\dagger_{tk'\sigma'}\, \hat{a}_{sk\sigma} \left( \frac{1}{2}\hat{\boldsymbol{\sigma}} \cdot \hat{\mathbf{S}} + U\hat{\sigma}_0 \right), \\
\hat{V}_{s \to s} &= \frac{1}{2} \sum_{\substack{\sigma,\sigma' \\ k,k'}} J_K\, \hat{a}^\dagger_{sk'\sigma'}\, \hat{a}_{sk\sigma}\hat{\boldsymbol{\sigma}} \cdot \hat{\mathbf{S}}.
\end{aligned} \tag{4}$$

Transition rates between the initial $|\Psi_\text{i}\rangle$ and final $|\Psi_\text{f}\rangle$ eigenstate of $\hat{H}_{\text{C}_{3v}}$ due to the interaction with electrons originating from the reservoir $j$ and absorbed in reservoir $j'$ are calculated using Fermi's golden rule:

$$\Gamma_{if}^{j \to j'}(eV) = \frac{G_{jj'}}{e^2} \int_{-\infty}^{\infty} d\epsilon\, W_{i \to f} f(\epsilon - eV, T_\text{eff})[1 - f(\epsilon - \epsilon_f + \epsilon_i, T_\text{eff})], \tag{5}$$

with $\epsilon_i$ and $\epsilon_f$ as the energy of the initial and final state, respectively, $f(\epsilon, T)$ as the Fermi-Dirac distribution, $T_\text{eff} \approx 0.65$ K as the effective temperature in the experiment, $G_\text{st} = G_\text{ts} = T_0^2 e^2$ adjusted to match the experimentally set conductance $I_\text{stab}/V_\text{stab}$, and $W_{i \to f}$ as the transition probabilities evaluated up to second order Born approximation:

$$W_{i \to f} = \frac{2\pi}{\hbar} \left( |M_{i \to f}|^2 + J_K \rho_\text{s} \sum_m \left( \frac{M_{i \to m} M_{m \to f} M_{f \to i}}{\epsilon_i - \epsilon_m} + \text{c. c.} \right) \right). \tag{6}$$

The matrix elements are evaluated as $M_{i \to j} = \sum_{i',j'} \sqrt{\lambda_{i'}\lambda_{j'}}\langle \varphi_{j'}, \Psi_j | \frac{1}{2}\hat{\boldsymbol{\sigma}} \cdot \hat{\boldsymbol{S}} + U\hat{\sigma}_0 | \varphi_{i'}, \Psi_i \rangle$ with $\varphi_{i',j'}$ as the eigenvectors and $\lambda_{i',j'}$ as the eigenvalues of the density matrices $\varrho_\text{t}$ and $\varrho_\text{s}$ of the electrons in tip and substrate participating in the scattering process.

The set of rate equations (Equ. 5) enables us now to build characteristic master equations for the state populations $p_i$ in which we take excitations and de-excitations of the spin system by the tunneling electrons and bath electrons into account [6,10,11]:

$$\frac{\mathrm{d}}{\mathrm{d}t} p_f(t) = \sum_{i \neq f} \sum_{j,j'} p_i(t) \Gamma_{if}^{j \to j'}(eV) - p_f(t) \sum_{i \neq f} \sum_{j,j'} \Gamma_{fi}^{j \to j'}(eV). \quad (7)$$

From the steady-state occupation $\bar{p}_i(eV)$ and the transition rates (equation 5) we can continue to calculate the current

$$I(eV) = e \sum_{i,j} \bar{p}_i(eV) \left( \Gamma_{ij}^{\mathrm{t} \to \mathrm{s}}(eV) - \Gamma_{ij}^{\mathrm{s} \to \mathrm{t}}(eV) \right), \quad (8)$$

and by numerical differentiation $\mathrm{d}I/\mathrm{d}V$. The individual state lifetime $\tau_i$ for the eigenstate $\Psi_i$ is the reciprocal of the sum of all scattering probabilities of events leaving the state:

$$\tau_i(eV) = \left( \sum_{f \neq i} \sum_{j,j'} \Gamma_{if}^{j \to j'}(eV) \right)^{-1}. \quad (9)$$

**Fitting details:** Due to the complex spectral features observed in the experimental data we refrained from using a fully automated least-square fitting procedure. Alternatively, we adjusted the parameters manually for each spectrum until a good agreement between simulation and experiment was found. In particular, for the sharp features at the transition between ground and first excited state at $V \approx \pm 1$ mV. By averaging the parameters found for all individual spectra, this approach enabled us to find a parameter set (Table 1 of the main text), which still reproduces all the experimental data very well (see Supplementary Figure 5 below). Note, that the potential scattering parameter given in all tables, $U = U_{\mathrm{pot}}/J_{\mathrm{K}}$, determines the ratio between potential scattering strength ($U_{\mathrm{pot}}$) and the Kondo scattering strength ($J_{\mathrm{K}}$). Magnitude and sign of this parameter strongly affect the overall spectral line shape. A positive $U$ describes an effective elastic scattering of electrons which possess the same spin direction as the atom, while for negative $U$ electrons with antiparallel spin have a larger elastic transition probability.

**Origin of the complex line shapes:** The origin of the observed complex line shapes of the SP-ISTS spectra is the joint effect of the current-induced reoccupation of the spin-states enabled by the large spin lifetimes and the non-zero spin-transfer torque caused by the current's spin-polarization. The intensities for elastic and also inelastic transitions are dependent on the actual spin-state of the atom but also on the spin of the tunneling electrons. E.g., a positive $U$ reflects a larger elastic tunneling for electrons whose spins are parallel to the spin of the atom, while a negative $U$ prefers tunneling of electrons with antiparallel spins. Thereby $U$ not only scales the zero-bias conductance in the absence of tunneling induced spin-flips but, in a spin-polarized measurement, also determines the overall line shape of the spectra in dependence on the actual state occupations. When $U$ is large, the upwards steps representing the spin-excitation in a spin-averaging ISTS measurement can hence be replaced by a strong drop of the differential conductance. In this case, the increased inelastic tunneling probability, which in a spin-averaged ISTS measurement leads to the upwards steps, is overcompensated by a strong reduction of the elastic transition probability.

Although spin-pumping is caused by inelastic transitions, the corresponding line shapes are hence strongly affected by the potential scattering. In contrast, in a spin-averaging measurement, elastic as well as inelastic contributions of spin-up and spin-down electrons cancel-out.

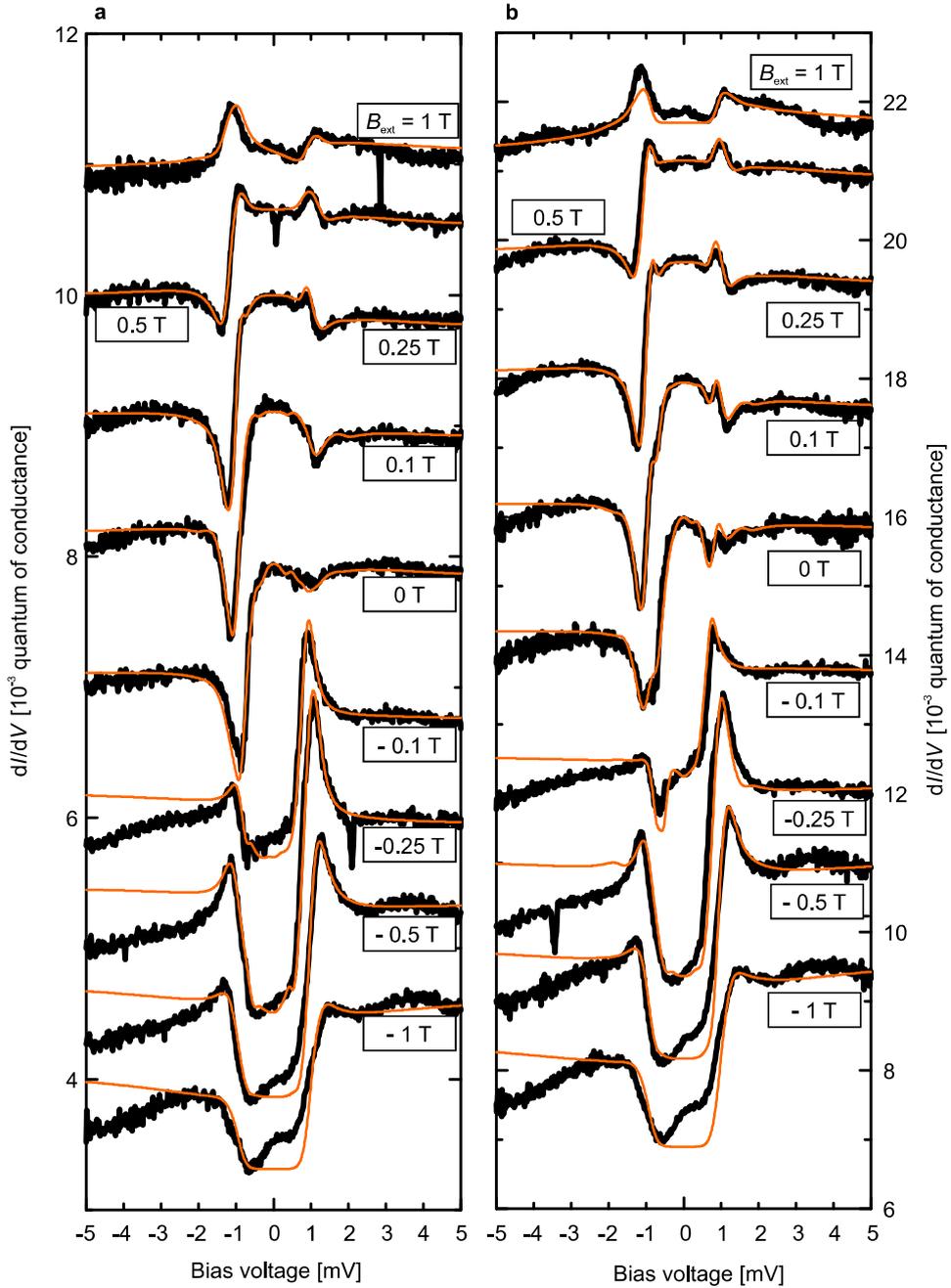

**Supplementary Figure 3** | Simulations (orange) of the experimental data (black) taken at the indicated external magnetic fields using a stabilization current of (**a**) $I_{\mathrm{stab}} = 3\,\mathrm{nA}$ and (**b**) $I_{\mathrm{stab}} = 6\,\mathrm{nA}$ ($V_{\mathrm{stab}} = 5\,\mathrm{mV}$, $V_{\mathrm{mod}} = 80\,\mathrm{\mu V}$).

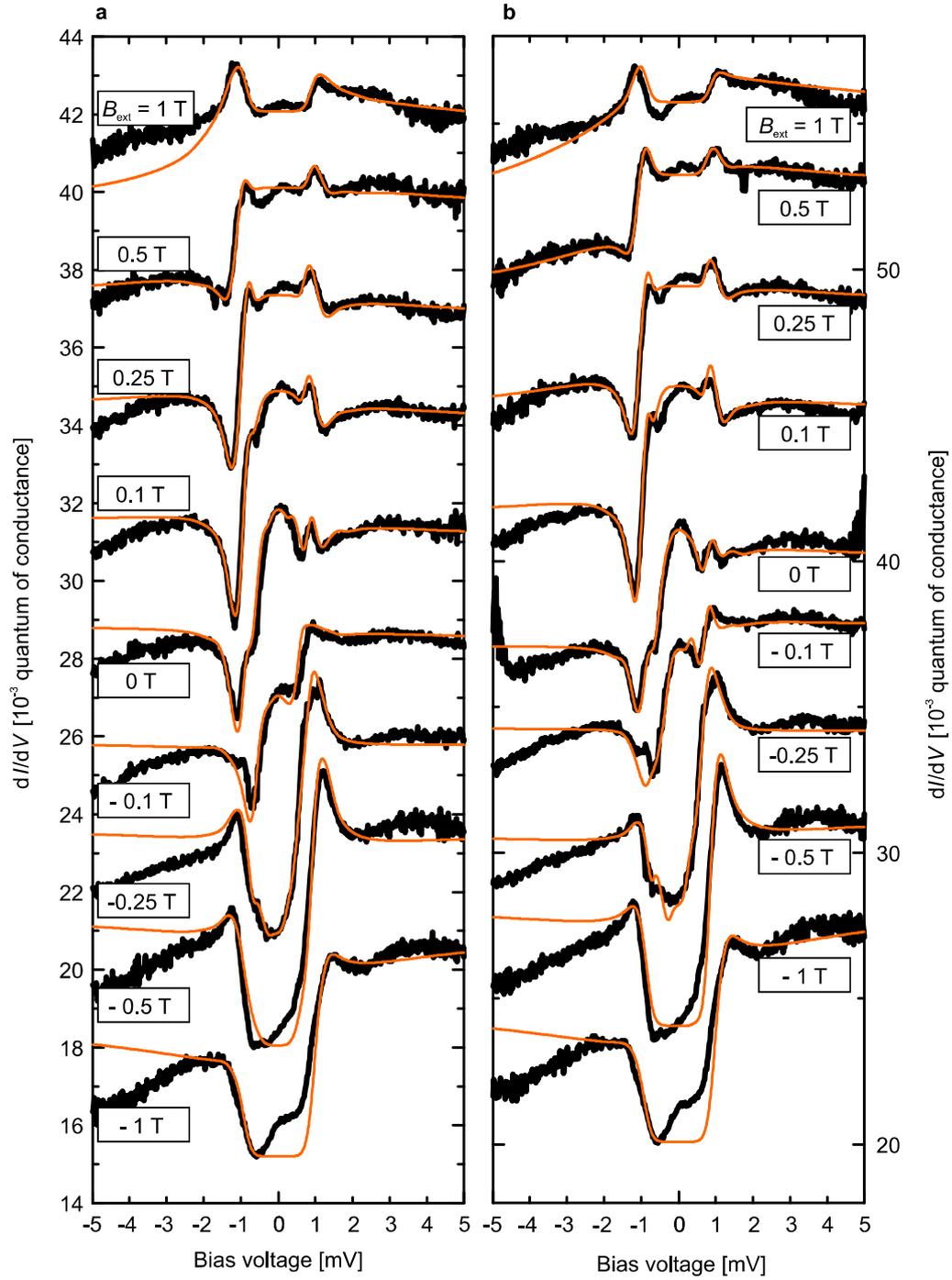

**Supplementary Figure 4** | Simulations (orange) of the experimental data (black) taken at the indicated external magnetic fields using a stabilization current of (**a**) $I_{\text{stab}} = 12\,\text{nA}$ and (**b**) $I_{\text{stab}} = 16\,\text{nA}$ ($V_{\text{stab}} = 5\,\text{mV}$, $V_{\text{mod}} = 80\,\mu\text{V}$).

# Supplementary Note 4 | Simulation parameters

| $I_\text{stab}, B_\text{ext}$ (nA),(T) | $B_2^0$ (meV) | $B_4^0$ (meV) | $B_4^3$ (meV) | $U$ | $J_\text{K}\rho_\text{s}$ | $T_0^2$ | $T_\text{eff}$ (K) | $\eta_t$ | $B_\text{sim}$ (T) | $B_\text{y}$ (T) |
|---|---|---|---|---|---|---|---|---|---|---|
| 1,-1    | -0.087 | 0.001  | 0      | 3.65 | 0.05       | 0.000085  | 0.7  | 0.17  | -1.1  | 0.5  |
| 1,-0.5  | -0.083 | 0.0018 | 0      | 3.53 | 0.04       | 0.000088  | 0.7  | 0.175 | -0.75 | 0.3  |
| 1,-0.25 | -0.083 | 0.0016 | 0      | 3.1  | 0.02       | 0.000107  | 0.6  | 0.145 | -0.4  | 0.25 |
| 1,-0.1  | -0.087 | 0.0012 | 0      | 4.1  | (+)0.015   | 0.0000675 | 0.55 | 0.16  | -0.35 | 0.2  |
| 1,0     | -0.083 | 0.001  | 0.0005 | 2.57 | 0.03       | 0.000142  | 0.6  | 0.145 | 0.4   | 0    |
| 1,+0.1  | -0.085 | 0.001  | 0      | 4.84 | 0.018      | 0.000048  | 0.6  | 0.215 | 0.65  | 0.2  |
| 1,+0.25 | -0.087 | 0.0015 | 0      | 4.82 | 0.03       | 0.000048  | 0.5  | 0.205 | 0.7   | 0.4  |
| 1,+0.5  | -0.09  | 0.0018 | 0      | 4.19 | (-)0.022   | 0.000063  | 0.7  | 0.15  | 0.95  | 0.2  |
| 1,+1    | -0.09  | 0.0018 | 0      | 4.22 | (-)0.024   | 0.000063  | 0.7  | 0.09  | 1.6   | 0    |
| 3,-1    | -0.083 | 0.0012 | 0      | 3.78 | 0.06       | 0.000245  | 0.7  | 0.165 | -1.2  | 0.5  |
| 3,-0.5  | -0.087 | 0.0013 | 0.0005 | 3.36 | 0.06       | 0.000293  | 0.7  | 0.17  | -0.8  | 0.3  |
| 3,-0.25 | -0.083 | 0.0016 | 0      | 3.11 | 0.04       | 0.000325  | 0.6  | 0.16  | -0.38 | 0.25 |
| 3,-0.1  | -0.087 | 0.0012 | 0.0005 | 4.93 | 0.04       | 0.000143  | 0.6  | 0.18  | -0.28 | 0    |
| 3,0     | -0.083 | 0.0012 | 0.003  | 2.55 | 0.07       | 0.00043   | 0.7  | 0.19  | 0.37  | 0.6  |
| 3,+0.1  | -0.09  | 0.001  | 0.003  | 4.63 | 0.055      | 0.000157  | 0.6  | 0.215 | 0.55  | 0.2  |
| 3,+0.25 | -0.087 | 0.0015 | 0.002  | 4.66 | 0.04       | 0.000155  | 0.5  | 0.215 | 0.7   | 0.4  |
| 3,+0.5  | -0.09  | 0.0018 | 0      | 4.2  | 0.05       | 0.000188  | 0.7  | 0.165 | 0.95  | 0.15 |
| 3,+1    | -0.09  | 0.0018 | 0.004  | 3.64 | (-)0.022   | 0.000248  | 0.7  | 0.07  | 1.6   | 0    |
| 6,-1    | -0.083 | 0.0012 | 0.001  | 3.4  | 0.13       | 0.00059   | 0.7  | 0.16  | -1.1  | 0.4  |
| 6,-0.5  | -0.083 | 0.0012 | 0.001  | 3.37 | 0.13       | 0.00059   | 0.7  | 0.17  | -0.75 | 0.4  |
| 6,-0.25 | -0.083 | 0.0012 | 0.003  | 3.3  | 0.11       | 0.00059   | 0.6  | 0.16  | -0.4  | 0.2  |
| 6,-0.1  | -0.083 | 0.0012 | 0.0015 | 3.69 | 0.09       | 0.00048   | 0.6  | 0.19  | -0.03 | 0.2  |
| 6,0     | -0.083 | 0.0015 | 0.0035 | 3.01 | 0.11       | 0.00067   | 0.7  | 0.195 | 0.25  | 0.2  |
| 6,+0.1  | -0.083 | 0.0015 | 0.0035 | 3.26 | 0.08       | 0.00058   | 0.55 | 0.19  | 0.5   | 0.2  |
| 6,+0.25 | -0.083 | 0.0015 | 0.0015 | 3.65 | 0.08       | 0.00048   | 0.55 | 0.18  | 0.7   | 0.2  |
| 6,+0.5  | -0.087 | 0.0015 | 0.0015 | 3.65 | 0.08       | 0.00048   | 0.55 | 0.165 | 0.95  | 0.3  |
| 6,+1    | -0.087 | 0.0015 | 0.003  | 3.68 | 0.08       | 0.00048   | 0.55 | 0.1   | 1.45  | 0.3  |

**Supplementary Table 1** | Parameters used for the simulations in Fig.3a of the main text as well as in Supplementary Fig.3. For simulations which are performed including second order scattering processes, the sign of $J_\text{K} \cdot \rho_\text{s}$ is given explicitly.

| $I_{\text{stab}}, B_{\text{ext}}$ (nA),(T) | $B_2^0$ (meV) | $B_4^0$ (meV) | $B_4^3$ (meV) | $U$ | $J_K\rho_s$ | $T_0^2$ | $T_{\text{eff}}$ (K) | $\eta_t$ | $B_{\text{sim}}$ (T) | $B_y$ (T) |
|---|---|---|---|---|---|---|---|---|---|---|
| 12,-1 | -0.083 | 0.0012 | 0 | 3.4 | 0.15 | 0.00118 | 0.7 | 0.16 | -1.2 | 0.5 |
| 12,-0.5 | -0.083 | 0.0017 | 0.0001 | 3.58 | 0.14 | 0.00105 | 0.7 | 0.17 | -0.6 | 0.5 |
| 12,-0.25 | -0.083 | 0.0016 | 0.0001 | 3.69 | 0.14 | 0.00097 | 0.7 | 0.18 | -0.2 | 0.4 |
| 12,-0.1 | -0.08 | 0.001 | 0.003 | 4 | 0.12 | 0.00083 | 0.7 | 0.21 | 0.045 | 0.4 |
| 12, 0 | -0.083 | 0.0015 | 0.004 | 3 | 0.11 | 0.00134 | 0.65 | 0.18 | 0.35 | 0.3 |
| 12, +0.1 | -0.083 | 0.0015 | 0.002 | 3.2 | 0.13 | 0.0012 | 0.7 | 0.185 | 0.55 | 0.3 |
| 12, +0.25 | -0.083 | 0.0016 | 0.001 | 3.19 | 0.15 | 0.0012 | 0.7 | 0.17 | 0.7 | 0.3 |
| 12, +0.5 | -0.087 | 0.0015 | 0.003 | 3.17 | 0.15 | 0.0012 | 0.7 | 0.17 | 0.9 | 0.4 |
| 12, +1 | -0.087 | 0.0015 | 0.003 | 3.17 | 0.15 | 0.0012 | 0.7 | 0.16 | 1.4 | 0.4 |
| 16, -1 | -0.083 | 0.0012 | 0 | 3.19 | 0.22 | 0.00177 | 0.7 | 0.16 | -1.2 | 0.4 |
| 16, -0.5 | -0.083 | 0.0012 | 0 | 3.58 | 0.22 | 0.0014 | 0.7 | 0.17 | -0.65 | 0.4 |
| 16, -0.25 | -0.083 | 0.0015 | 0 | 3.75 | 0.22 | 0.00125 | 0.7 | 0.2 | -0.12 | 0.4 |
| 16, -0.1 | -0.077 | 0.0014 | 0.002 | 3.48 | 0.2 | 0.0014 | 0.7 | 0.195 | 0.11 | 0.3 |
| 16, 0 | -0.074 | 0.00155 | 0.005 | 3 | 0.14 | 0.00173 | 0.65 | 0.185 | 0.35 | 0.35 |
| 16, +0.1 | -0.08 | 0.0016 | 0.0015 | 3.42 | 0.22 | 0.00141 | 0.6 | 0.19 | 0.55 | 0.25 |
| 16, +0.25 | -0.083 | 0.0017 | 0.005 | 3.81 | 0.22 | 0.00116 | 0.7 | 0.185 | 0.7 | 0.1 |
| 16, +0.5 | -0.083 | 0.0016 | 0.0005 | 3.8 | 0.23 | 0.00116 | 0.7 | 0.185 | 0.95 | 0.1 |
| 16, +1 | -0.083 | 0.0014 | 0.0005 | 3.8 | 0.22 | 0.00116 | 0.7 | 0.185 | 1.5 | 0.1 |

**Supplementary Table 2** | Parameters used for the simulations in Supplementary Fig.4.

# Supplementary Note 5 | Simulations using the averaged parameter set

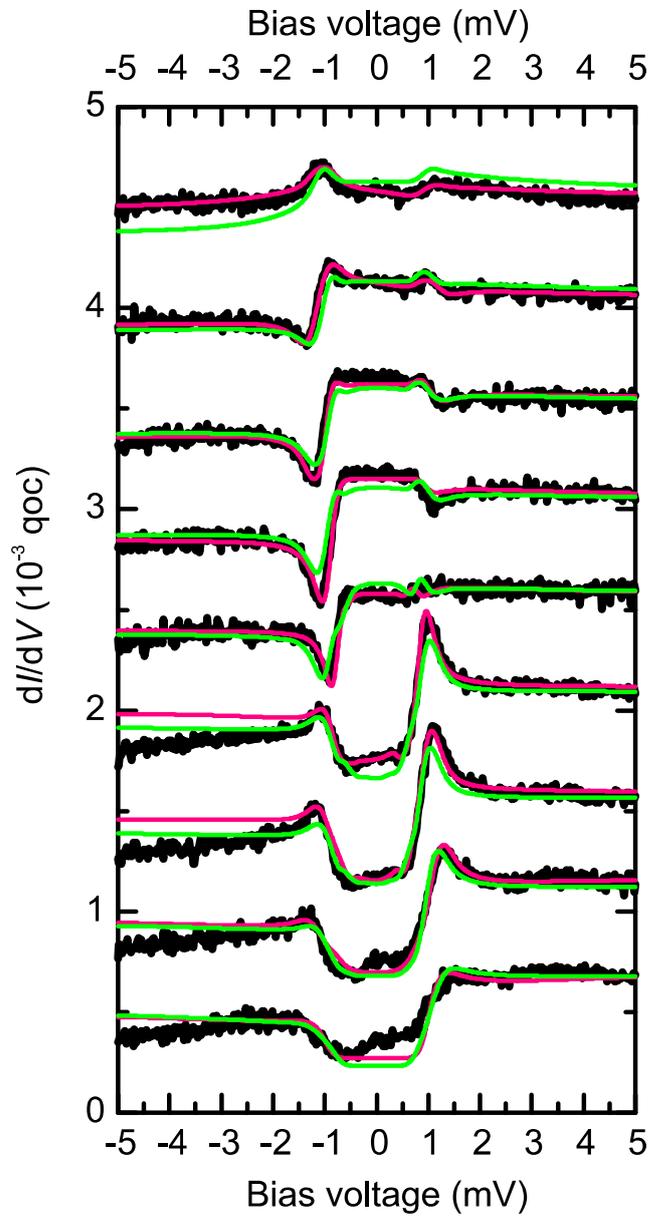

**Supplementary Figure 5** | Comparison of the experimental data of Fig.3a of the main text with the simulation using the averaged parameter set given in Table 1 of the main text (green), and with the simulation obtained from the individually optimized parameters given in Supplementary Table 1 (red).

In Fig.3a of the main text a very good agreement between the individually optimized simulations and the experimental data was shown, resulting in the parameters given in Supplementary Table 1. In order to check the validity of the averaged simulation parameters given in Table 1 of the main text, Supplementary Fig.5 shows a comparison of the data with the simulation using the individually optimized parameters (Supplementary Table 1), and the simulation using the averaged parameters (Table 1 of the main text). Note, that for $J_{\mathrm{K}} \cdot \rho_{\mathrm{s}}$, the averaging was restricted to the values from the 1 nA simulations, and for the out-of-plane as well as the in-plane magnetic fields, the values from Supplementary Table 1 were used. Therefore, the only free simulation parameter was $T_0^2$. Obviously, the main features of the SP-ISTS data are already well-described by the set of averaged parameters, which corroborates their validity.

**Supplementary References**